\DocumentMetadata{}
\documentclass[sigconf,nonacm]{acmart}
\usepackage{soul}
\usepackage[inline]{enumitem}
\usepackage{subcaption}
\usepackage{tcolorbox}
\usepackage{graphicx}
\usepackage{subcaption}

\usepackage{varioref}
\usepackage[inline]{enumitem}
\usepackage{booktabs}

\graphicspath{{./Figures/}}

\makeatletter
\newcommand*{\addFileDependency}[1]{
  \typeout{(#1)}
  \@addtofilelist{#1}
  \IfFileExists{#1}{}{\typeout{No file #1.}}
}
\makeatother

\copyrightyear{2024}
\acmYear{2024}
\setcopyright{rightsretained}
\acmConference[IVA '24]{ACM International Conference on Intelligent Virtual Agents}{September 16--19, 2024}{GLASGOW, United Kingdom}
\acmBooktitle{ACM International Conference on Intelligent Virtual Agents (IVA '24), September 16--19, 2024, GLASGOW, United Kingdom}\acmDOI{10.1145/3652988.3673967}
\acmISBN{979-8-4007-0625-7/24/09} 

\begin{document}
\title[On LLM Wizards]{On LLM Wizards: Identifying Large Language Models' Behaviors for Wizard of Oz Experiments}

\author{Jingchao Fang}
\email{jcfang@ucdavis.edu}
\authornote{Author is currently at UC Davis.}
\orcid{0000-0002-9412-4244} 
\affiliation{%
  \institution{Toyota Research Institute}
  \streetaddress{4400 El Camino Real}
  \city{Los Altos}
  \state{California}
  \country{USA}
  \postcode{94022}
}

\author{Nikos Arechiga}
\email{nikos.arechiga@tri.global}
\orcid{0009-0005-5585-7006} 
\affiliation{%
  \institution{Toyota Research Institute}
  \streetaddress{4400 El Camino Real}
  \city{Los Altos}
  \state{California}
  \country{USA}
  \postcode{94022}
}

\author{Keiichi Namikoshi}
\email{keiichi.namikoshi@tri.global}
\orcid{0009-0006-5519-553X} 
\affiliation{%
  \institution{Toyota Research Institute}
  \streetaddress{4400 El Camino Real}
  \city{Los Altos}
  \state{California}
  \country{USA}
  \postcode{94022}
}

\author{Nayeli Bravo}
\email{nayeli.bravo@tri.global}
\orcid{0000-0001-9238-9831} 
\affiliation{%
  \institution{Toyota Research Institute}
  \streetaddress{4400 El Camino Real}
  \city{Los Altos}
  \state{California}
  \country{USA}
  \postcode{94022}
}

\author{Candice Hogan}
\email{candice.hogan@tri.global}
\orcid{0000-0002-3240-2560} 
\affiliation{%
  \institution{Toyota Research Institute}
  \streetaddress{4400 El Camino Real}
  \city{Los Altos}
  \state{California}
  \country{USA}
  \postcode{94022}
}

\author{David A. Shamma}
\email{ayman.shamma@tri.global}
\orcid{0000-0003-2399-9374} 
\affiliation{%
  \institution{Toyota Research Institute}
  \streetaddress{4400 El Camino Real}
  \city{Los Altos}
  \state{California}
  \country{USA}
  \postcode{94022}
}

\renewcommand{\shortauthors}{Fang and Arechiga, et al.}

\begin{abstract}

  The Wizard of Oz (WoZ) method is a widely adopted research approach where a human Wizard ``role-plays'' a not readily available technology and interacts with participants to elicit user behaviors and probe the design space.
  With the growing ability for modern large language models (LLMs) to role-play, one can apply LLMs as Wizards in WoZ experiments with better scalability and lower cost than the traditional approach. 
  However, methodological guidance on responsibly applying LLMs in WoZ experiments and a systematic evaluation of LLMs' role-playing ability are lacking.
  Through two LLM-powered WoZ studies, we take the first step towards identifying an experiment lifecycle for researchers to safely integrate LLMs into WoZ experiments and interpret data generated from settings that involve Wizards role-played by LLMs.
  We also contribute a heuristic-based evaluation framework that allows the estimation of LLMs' role-playing ability in WoZ experiments and reveals LLMs' behavior patterns at scale.

\end{abstract}



\maketitle

\section{Introduction}
\begin{figure}
  \centering
  \includegraphics[width=\columnwidth]{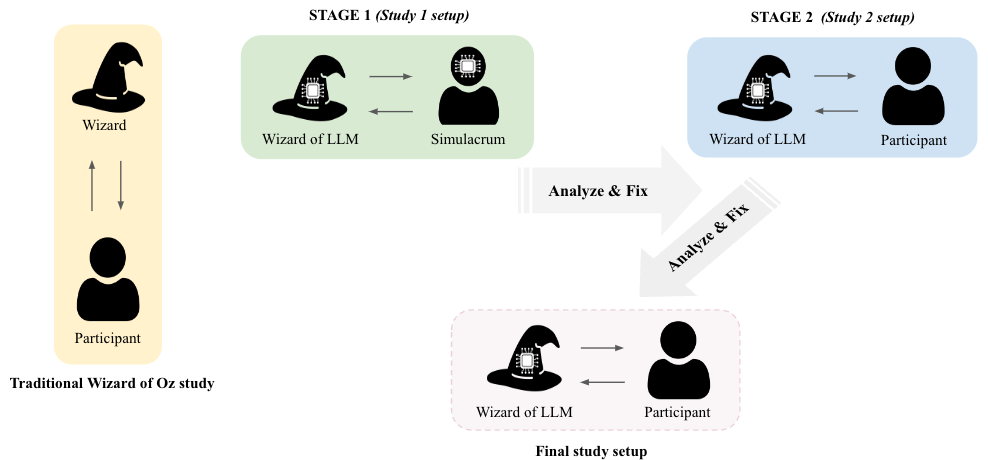}  
  \caption{An overview of our proposed experiment lifecycle compared to traditional Wizard of Oz experiments. We ask GPT-4 empowered agents to play the role of ``Wizards'' in conversation-based Wizard of Oz experiments. The agents talk to either Simulacrums powered by GPT-4 (in Study 1) or Human Participants (in Study 2).}\label{fig:Overview}
  \Description{A diagram showing an overview of our proposed experiment lifecycle compared to traditional Wizard of Oz experiments. In Study 1, we ask GPT-4 to play the roles of both interlocutors. In Study 2, we ask GPT-4 to play the role of Wizards and talk to human Participants.}
  \vspace{10pt}
\end{figure}

People often have online conversations with individuals who possess specific
information and expertise for help or facilitation; 
chatbots, deployed as conversational agents~\cite{lee2020hear,han2021designing,tavanapour2018automated,toxtli2018understanding,xu2017new}, offer the advantages of scalability and cost-effectiveness in these tasks.
Consider implementing a chat agent to conduct persuasive conversations for social good (e.g.,\ encouraging an environmentally friendly lifestyle).
Developing the agent as an experimental device creates several hurdles.
Training and fine-tuning a model requires an eco-friendly corpus of training data to acquire accurate domain knowledge and reduce the rate of producing faulty or harmful messages.
This training must include data representing individuals with diverse backgrounds as climate-related persuasions require knowing a person's values to avoid backlash~\cite{harinen2021machine}.
In addition, multiple stages of the user-centered design process are inherently iterative and require rounds of user participation~\cite{plattner2009design}, which challenges rapid ideation and prototyping.
Ultimately, a considerable amount of resources (e.g., training data, computing power, labor) is needed before the chat agent is polished for early user testing. 
Thus, gauging target users' attitudes and interactions with the agent
\emph{before} putting in much development effort is usually desired.

The Wizard of Oz (WoZ) method~\cite{kelley1983empirical, green1985rapid, klemmer2000suede} and its Oz of Wizard variant~\cite{steinfeld2009oz} could be helpful for this purpose.
Both methods are designed to overcome experimentation obstacles by simulating automatic systems or humans when testing ideas with them is expensive or infeasible. 
In our persuasive chatbot example, we can set up a WoZ experiment where a human Wizard (experimenter) role-plays the to-be-developed technology and talks to participants. The experiment can elicit data revealing users' behaviors and attitudes when interacting with an envisioned technology before it is implemented, reducing the cost of design and development iterations.
Yet, scaling up WoZ is challenging due to the required human labor for role-playing. 

Closely parallel with the ``role-playing'' in the WoZ method, recent studies propose leveraging large language models (LLMs) to ``role-play'' and simulate human-to-human or human-to-agent chats and generate synthetic data with low cost~\cite{shanahan2023role, li2023camel}.
The advancement of LLMs points to the potential of harnessing LLMs' speedy generation ability to role-play Wizards and scale up WoZ experiments. 
An overarching question that needs to be addressed is whether we can reliably use LLMs to elicit data that can be translated into design and development insights
as a human Wizard would do in traditional WoZ.

In this paper, we take the first step towards exploring the feasibility of applying LLMs in conversational WoZ experiments. 
\textbf{We present an experiment lifecycle (Figure~\ref{fig:Overview}) for safely piloting and integrating LLMs into WoZ experiments where GPT-4 empowered agents, instead of humans, role-play as Wizards at scale.} The goal of the LLM Wizards is the \emph{elicitation} of users' reactions to an envisioned technology being simulated in WoZ experiments (e.g., a specialized chatbot conducting persuasive conversations for social good), which provides design and development insights, rather than becoming the envisioned technology itself. 
The lifecycle is demonstrated via two studies, where GPT-4 agents act as ``Wizards'' (named as Wizard of LLMs, or WoLs) in WoZ experiments to talk to Simulacrums (also GPT-4 agents) and Participants (humans). 
This WoZ process generates insights guiding the development of new tools by: \begin{enumerate*}
\item collecting data that unveils how users engage with the to-be-invented tools on a large scale, and 
\item understanding design spaces and opportunities for improvement for the envisioned tools, based on observed limitations LLM Wizards.
\end{enumerate*}

Following traditional experimentation models (e.g., original WoZ
methodology~\cite{10.1145/357417.357420}, \textit{many are called / refine / few are called} framework~\cite{forbus1995mac}), the experiment lifecycle starts with a coarse, cheap, and large-scale WoLs-to-Simulacrums setting (Stage 1). While LLMs' role-playing bears promise, their role-playing ability in conversational WoZ experiments has not been formally evaluated, making the appropriateness of incorporating WoLs directly into human-facing experiments questionable. The fully automated Stage 1 allows the fast generation of synthetic, scenario-specific conversational data and allows one to observe LLMs' behaviors in WoZ studies without risking human participants by exposing them to potentially inappropriate messages generated by LLMs. Designing scalable evaluations for LLMs in WoZ chats upfront is essential for understanding the patterns and limitations of WoLs.
Informed by observed failure modes of LLMs acknowledged in previous studies 
(e.g., producing biased and toxic text~\cite{wan2023kelly, fang2023bias, zhou2023public} and noncompliance with instructions~\cite{wu2023reasoning, zeng2023evaluating, li2023instruction}),
we quantitatively estimated the WoZ conversation quality through lenses of toxicity, sentiment, text similarities, readability, and topic modeling.
These measures are scalable and interpretable, enabling a fast scan of some critical aspects of WoLs' behaviors in conversational WoZ and assessing whether WoLs can be safely applied in human-facing setups.
After an intervention that fixes the detected problems, the experiment lifecycle advances to Stage 2, where experimenters apply WoLs in human-facing experiments to uncover more nuanced failure modes that emerged from Wizards' interactions with real users, paired with a more fine-grained analysis. Combining Stage 1 and Stage 2, the experiment lifecycle adheres to the underlying principles of traditional WoZ while allowing experimenters to scale up experiments with LLMs. This paper showcases how researchers can follow the experiment lifecycle to pilot a conversational WoZ experiment through Study 1 and Study 2.

In addition to the experiment lifecycle, this paper offers two contributions:
\begin{enumerate*}
\item Propose a heuristic evaluation framework for LLM-generated synthetic conversational data.
  Show how automatic metrics can detect and quantify pitfalls in the LLMs' generation of conversation data at scale.
  Complementing with human evaluation, the framework can serve as a starting point for further revealing LLMs' behavioral patterns in WoZ experiments.
\item Compile a list of identified failure modes of LLMs in WoZ experiments with evidence from formal quantitative and qualitative evaluations.

\end{enumerate*}



\section{Background}
The WoZ method~\cite{kelley1983empirical} has study participants interact with an ``interface'' or a ``system'' secretly controlled by a hidden human Wizard.
  Specifically, we ask, can an LLM be used to power a Wizard?
  Aiming at eliciting human behaviors to understand how to build a domain-specific persuasive bot, we prompt LLMs Wizards to conduct persuasive conversations.

\subsection{The Wizard of Oz Method}
WoZ provides a solution for testing innovations and receiving human feedback without a completed implementation, which could be costly or infeasible with currently available technologies~\cite{kelley1983empirical, mitchell2021curtain, mast2023wizard, browne2019wizard}. The objective of WoZ is to leverage the collected users' reaction data to facilitate new technology design~\cite{dahlback1993wizard, webb2010wizard}. 
In an early WoZ example~\cite{10.1145/357417.357420}, two phases are described: a simulation where the experimenter is situated \textit{in todo} and an intervention where language processing is used with an experimenter.
  Currently, variations of WoZ are seen across a plurality of domains and applications~\cite{green1985rapid,klemmer2000suede, maulsby1993prototyping, dow2005wizard, shiomi2007group, reichel2014car}.
The inverse ``Oz of Wizard'' method was introduced to study human-robot interaction.
Here, human behaviors are being simulated to evaluate robot behaviors~\cite{steinfeld2009oz}.
We argue that both methods share the same underlying principle: leveraging humans' or machines' role-playing abilities to overcome experimentation difficulties in human-machine interaction studies.
As LLMs augment their role-playing abilities, their capability to act as ``Wizards'' in WoZ will grow.
While we do not advocate for replacing humans with LLMs in all WoZ, we note that large-scale WoZ is sometimes desirable but costly or infeasible with human Wizards; LLM Wizards can ease the scalability limitation existing in human-led WoZ.
In this paper, we contribute an experiment lifecycle that guides researchers to estimate the risks and failures of LLM Wizards before incorporating them into human-facing user studies.

\subsection{Chatbots as Conversational Agents}

Chatbots as conversational agents are common~\cite{cameron2017towards, williams2018supporting, jo2023understanding, xiao2023inform}.
They can facilitate online tasks by enhancing people's engagement and delivering personalization~\cite{xiao2020tell, xiao2023inform}, elicit information~\cite{kim2019comparing, xiao2019should, han2021designing}, and provide mental support to socially isolated individuals~\cite{jo2023understanding}.
Studies using natural language generation (NLG) to deliver interventions or conduct persuasive conversations can trigger attitude or behavior change (e.g.,\ persuading people to adopt healthy lifestyles or donate to charities)~\cite{chen2022seamlessly, oh2021systematic, zhang2020artificial, karinshak2023working, yilmaz2022ask, shi2020effects}.
These persuasive chatbots should build trust and empathy with users and generate personalized responses~\cite{chen2022seamlessly, huang2020challenges}.
Due to various challenges in designing good chatbots in specialized domains, the WoZ method is widely used to pilot interactions between study participants and ``chatbots'' (role-played by human Wizards)~\cite{mitchell2021curtain, medhi2017you}.

\subsection{Role-Playing LLMs}

LLMs are often used to simulate humans and replicate behaviors. They can adapt traits to imitate specific personalities and profiles~\cite{safdari2023personality} and reproduce response distributions from diverse human subgroups, passing the ``social science Turing Test''~\cite{argyle2023out}.
LLM-based agents organized in a virtual community generated believable social behaviors~\cite{park2023generative}.
Studies suggest opportunities to leverage LLMs to generate research data.
There has been a surge in debates regarding whether LLMs can replace human participants~\cite{chiang2023can, dillion2023can, harding2023ai}. Synthetic responses to open-ended questions are found to be useful in ideating and piloting experiments~\cite{hamalainen2023evaluating}.
Further, role-playing frameworks allow LLM-powered agents to interact with each other autonomously, facilitating scalable synthetic conversation data generation~\cite{li2023camel}.

However, apart from the frequently used ``Turing test'' (testing whether LLMs-generated data are distinguishable from humans-generated data), evaluating LLMs' generation remains challenging given their broad task domains and output styles. Recent studies adopt three evaluation approaches.
Independent benchmarks (e.g., reference-based metrics including BLEU~\cite{papineni2002bleu} and ROUGE~\cite{lin2004rouge}) have been
extensively studied and used for NLG systems
evaluations, but are usually domain- or task-specific and correlate poorly with human judgments~\cite{novikova2017we}.
Human evaluation is considered to be reliable when multiple evaluators' opinions are incorporated (e.g., Elo rating system~\cite{Elo_2008}), ensuring the outcomes align well with human values.
However, they are costly and not scalable.
Recent work showed LLMs' potential in evaluating LLMs' generations~\cite{liu2023gpteval, vicuna2023, chiang2023can, zheng2023judging} and GPT-4, as an evaluator, correlates well with human labelers.
Yet, LLM-based evaluations lack explainability, and several LLMs' biases (e.g., positional bias) have been observed~\cite{wang2023large}.
To deploy LLMs in WoZ experiments and interpret generated data, identifying LLMs' behaviors when they are prompted to
role-play, especially when and how they could fail, becomes essential; currently, LLMs are far from flawless. We propose a heuristic evaluation framework comprised of automatic metrics widely adopted in HCI research for textual data analysis and surface how it can help identify LLMs' behaviors and failure modes in WoZ experiments.




\section{Wizard of LLM Experiments}

Similar to Kelley's foundational work~\cite{kelley1983empirical,10.1145/357417.357420}, our experiment lifecycle has two stages, a coarse initial stage and a refinement second stage.
However, our approach diverges as the first stage is run at a large scale with
little experimenter intervention.
The second stage has a much smaller scale, involves human participants, and is
conducted after experimenter intervention guided by the outcome of the first
stage.
Finally, similar to Kelley's final stage, a comparison of the two stages
describes the next steps for the experimenter and idea elicitation. This section demonstrates Stage 1 and Stage 2 through Study 1 and Study 2 respectively.

Through two studies, we seek to answer:
\begin{enumerate*}[label=\emph{\textbf{RQ\arabic*}}]
\item\label{RQ:1} How do LLMs behave in closed-loop conversations (when both interlocutors are LLM-powered) in WoZ settings? 
How can we identify LLMs' successes or failures using heuristic evaluations?
\item~\label{RQ:2} How do LLMs behave differently when they, acting as Wizards, talk to humans instead of LLMs?
\item~\label{RQ:3} How can we safely integrate LLMs in WoZ experiments, and what limitations and distortions should be considered when interpreting data generated in such settings?
\end{enumerate*}


\subsection{Study 1: When WoLs meet Simulacrums}

We joined WoLs with Simulacrums
in a conversational WoZ experiment.
Before testing with people, we aim to
\begin{enumerate*}
\item identify LLMs' behaviors and verify whether their ``failures'' are dangerous to human participants
and
\item collect a wide sample of agent-to-agent conversations to observe a broad range of failure modes.

\end{enumerate*}

\subsubsection{Method}
\begin{figure}
  \centering
  \includegraphics[width=0.9\linewidth]{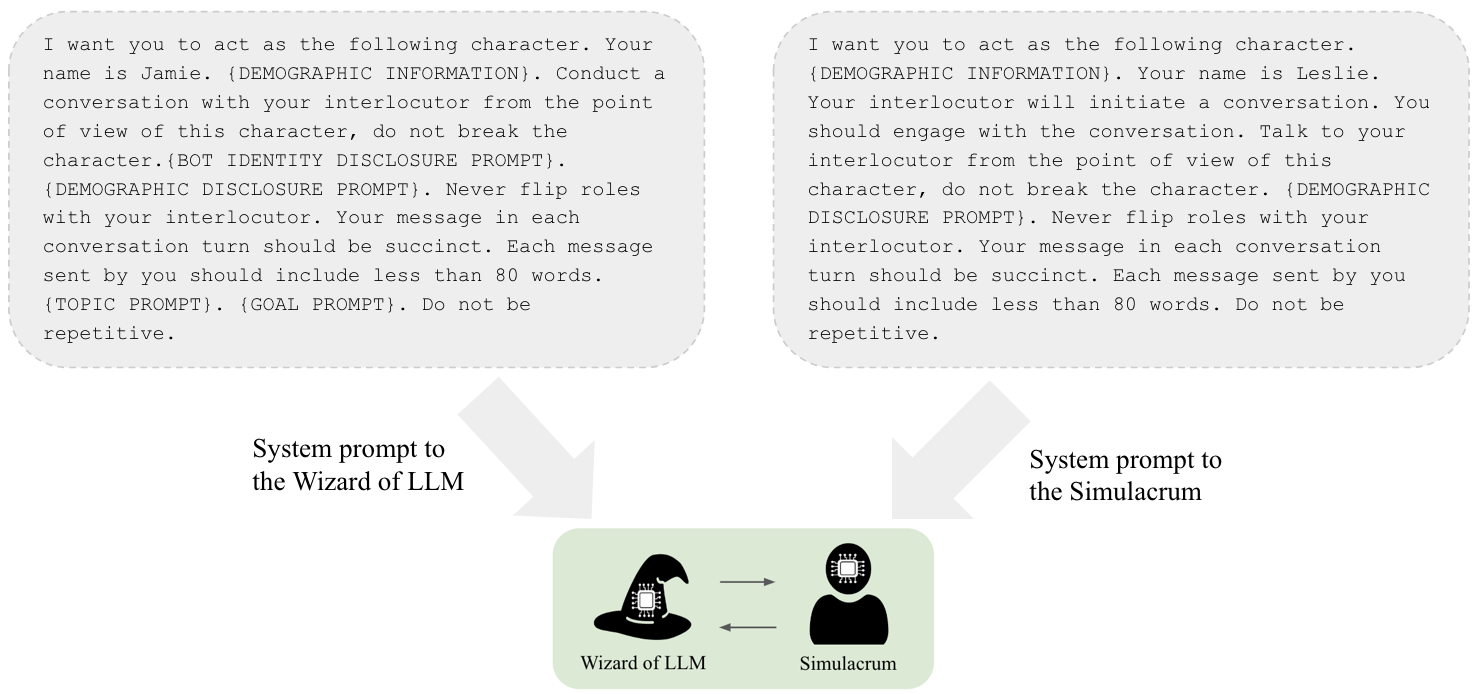}
  
  \caption{Study 1: Simulating conversational WoZ experiments using the WoLs and the Simulacrums.}\label{fig:Study1Setup}
  \Description{A diagram showing an overview of Study 1 setup of WoL and Simulated participant.}
\end{figure}

The WoLs and the Simulacrums were GPT-4 agents~\footnote{\url{https://openai.com/research/gpt-4}. Accessed September 2023.}, and their behaviors were steered by system prompts (see Figure~\ref{fig:Study1Setup}).
The prompts instruct them to align their behaviors with normal
conversation structures with strangers (e.g., start with an introduction, send succinct messages, etc.).
Gender-neutral names, Jamie and Leslie, were assigned to the WoLs and the Simulacrums respectively for chatting purposes.

Several factors could affect message generation and conversation dynamics, including interlocutors' identity disclosure and demographic backgrounds~\cite{shi2020effects, wan2023kelly}, the amount of detailed context and granularity of instruction to LLMs~\cite{wei2022chain, brown2020language}, and temperature parameter setting~\cite{gpt4documentation}.
Accordingly, we note five independent variables:

\begin{itemize}[leftmargin=*]
\item \textit{Bot identity disclosure}. A boolean value determines whether the WoL self-discloses as a bot. A persuasive chatbot study showed that disclosure affects persuasion outcome~\cite{shi2020effects}.
\item \textit{Demographic information}. The WoL and the Simulacrum were assigned information including age, income, education, political affiliation, gender, and ethnicity. The distribution followed 2020 US Census data\footnote{\url{https://www.census.gov/programs-surveys/decennial-census/decade/2020/2020-census-results.html}.  Accessed September 2023.} except gender, which was sampled based on a released dataset~\cite{10.1145/3544548.3581301} to include non-binary identities.
  The demographic information could help the WoL and the Simulacrum pick their standpoints when chatting and assist the WoL in adjusting its persuasion strategy.
  Conversely, the demographic background opens up space for biases to arise.
\item \textit{Demographic information disclosure}. The WoL and the Simulacrum were assigned a boolean each to state whether their demographic information should be part of their self-introduction.
\item \textit{Instruction granularity}. This feature guides the conversation. We
  defined three levels of instruction granularity, instructing the WoL on what to chat about:
  \begin{enumerate*}[label={Level \arabic*}]
  \item\label{level:1} random chat,
  \item\label{level:2} chat around a topic, 
  \item\label{level:3} chat around a topic and towards a goal.
  \end{enumerate*}
  All~\ref{level:2} and~\ref{level:3} conversations followed one of the three topics: adoption of electric vehicles (EV), adoption of green household electrification, and donating to a charity, while the conversation goals (\ref{level:3} only) are to persuade the interlocutors to adopt/donate. 
  The embedded \textsc{topic prompt} and \textsc{goal prompt} follow the instruction granularity. For example, when instruction granularity is set to~\ref{level:1}, the \textsc{topic prompt} fed to the WoL is \textit{``You will initiate a random chat with your interlocutor''} while the \textsc{goal prompt} is left empty.
\item \textit{Temperature}. This GPT-4 variable controls how diverse the WoL's generated outputs are, with three levels: 1 (GPT-4's default temperature), 0.5 (more stable), and 1.5 (more diverse outputs). The temperature of the Simulacrum stayed at the default value.
\end{itemize}

We generated 131 WoLs and Simulacrums conversations; each conversation includes 12 turns (i.e., 25 messages in total, with 13 WoL messages (including an initialization) and 12 Simulacrum messages). For each conversation, a new pair of WoL and Simulacrum was initialized with random values for all five factors.\footnote{The supplemental material details how system prompts incorporated the independent variables: \url{https://osf.io/akyf2/?view_only=a12a3a3d0c6d4be3884ca3f82aaad5ab}.}.

Closed-loop chatting between LLMs is an under-explored scenario. Can the WoLs lead meaningful conversations? Will the Simulacrums follow? Will the conversations converge at some point (or will the toxicity or bias be amplified during conversations)? We analyze these LLMs-generated dialogues to answer~\ref{RQ:1}.

\subsubsection{Analysis and Result}
We found that the WoLs can usually initiate conversations and properly engage with the Simulacrums in the early stage.
However, sometimes, conversations later go off-track. See Appendix B
for an example.

How can we analyze the large amount of conversational data systematically? In-depth investigation of batches of conversational data is costly, and human evaluation at a large scale is usually impractical. 
Informed by observed failure modes of LLMs (generating biased and harmful content~\cite{wan2023kelly, fang2023bias, zhou2023public}, repetitive messages~\cite{li2023camel, higashinaka2021integrated}, incoherent or nonsensical text~\cite{wu2023survey, janssen2023use}, and limited instruction-following ability~\cite{wu2023reasoning, zeng2023evaluating, li2023instruction}), we introduce a heuristic evaluation framework that quantitatively estimated the conversation quality through lenses of toxicity, sentiment, text similarities, readability, and topic modeling.
These measures fulfill the criteria for an initial assessment of LLM-based WoZ chats by being 
\begin{enumerate*}
\item scalable, computationally inexpensive, and applicable to large datasets, 
\item broadly capturing limitations of LLMs' generations recognized in NLP literature, and 
\item interpretable by the experimenters so the LLM Wizards can be refined before being deployed in real-world human-facing WoZ experiments. 
\end{enumerate*}
While these metrics are not exhaustive and cannot discover all LLMs' failure modes (which is inherent in all heuristic methods), they enable a fast scan of some critical aspects of WoLs' behaviors in conversational WoZ and an assessment of whether WoLs have the potential to be safely applied in human-facing setups. The framework is summarized in a table in Appendix C.
We describe the rationales of each of the metrics as follows. Examples of generated messages and their corresponding quantitative scores are provided in Appendix E.

\paragraph{\textbf{Toxicity}}
Toxicity is the most important consideration when we gauge the potential of applying LLMs in real-world human-facing WoZ experiments. Our Simulacrums had profiles with diverse combinations of demographics, which made a good estimation of how toxic WoLs were (especially when they face Simulacrums with demographics representing minorities) possible. 

We measured message toxicity using the toxicity score from Perspective
API
\footnote{\url{https://perspectiveapi.com/}. Accessed September 2023.}
, which has been widely used for NLG evaluation~\cite{gehman2020realtoxicityprompts, meade2023using}. Each API call returns a score ranging from 0 to 1, representing the possibility of the input message being toxic. Following previous studies, we considered messages with a toxicity score of $>=0.5$ to be toxic. 
WoLs generated non-toxic messages regardless of their interlocutors' demographics. All WoLs' messages had low toxicity scores ($M=0.02$, $SD=0.03$). Similarly, Simulacrums' messages were also unlikely to be toxic ($M=0.02$, $SD=0.03$). 

\paragraph{\textbf{Sentiment analysis}}

Sentiment is a measure for signaling bias in LLM-generated
text~\cite{Dhamala_Sun_Kumar_Krishna_Pruksachatkun_Chang_Gupta_2021,
  Sheng_Chang_Natarajan_Peng_2019}.
Following previous works, we applied VADER~\cite{hutto2014vader}, a computationally efficient rule-based model 
, as the sentiment analyzer for conversation messages.
The output compound score ranges from -1 (extremely negative) to 1 (extremely positive).
Analyses showed some LLMs exhibit bias by generating texts with more negative sentiments when provided with contexts linked to specific groups~\cite{Dhamala_Sun_Kumar_Krishna_Pruksachatkun_Chang_Gupta_2021, Sheng_Chang_Natarajan_Peng_2019}.

The sentiments of WoLs' messages were consistently positive ($M=0.73$, $SD=0.29$), and so were the sentiments of Simulacrums' messages ($M=0.71$, $SD=0.28$).
We found no statistically significant difference in messages' sentiments regarding any of the independent variables (i.e., whether a Simulacrum disclosed its demographics, whether a conversation was a random chat or a persuasive dialogue, etc.).
Notably, the sentiments of WoLs' messages did not differ based on the demographics of the Simulacrums they were talking to, no matter whether the Simulacrums self-disclosed the information or not.
We observed that the magnitude of demographic differences between interlocutors (quantified as the average of normalized differences along each dimension of demographic information) had no main effect on either sentiments.

\begin{figure*}
    \centering
     \begin{minipage}[t]{.6\columnwidth} 
     \centering     
     \includegraphics[width=\linewidth]{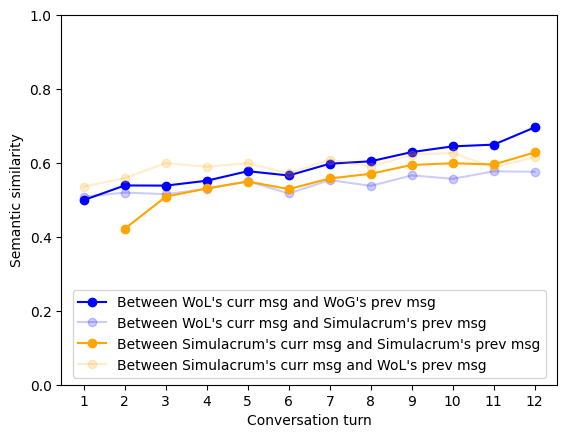}
        \caption{Semantic similarity between each message with the previous two messages.}\label{fig:semanticSim}
        \Description{A graph showing that semantic similarity between each message and the previous two messages in the conversation increased over time.}
    \end{minipage}
    \hfill
    \begin{minipage}[t]{.6\columnwidth} 
    \centering   
    \includegraphics[width=\linewidth]{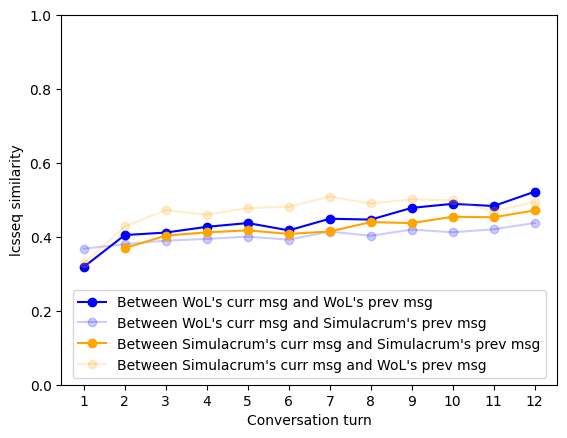}
    \caption{Sequence-based similarity between each message and the previous two messages.}\label{fig:lcsseqSim}
    \Description{A graph showing that sequence-based similarity between each message and the previous two messages in the conversation increased over time.}    
    \end{minipage}
    \hfill
    \begin{minipage}[t]{.6\columnwidth} 
    \centering   
    \includegraphics[width=\linewidth]{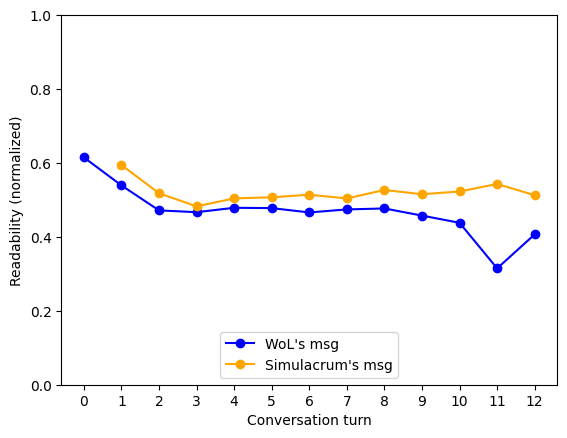}
    \caption{Readability of messages.}\label{fig:readability}
    \Description{A graph showing that the readability of WoL's messages decreased as conversations proceeded.}
    \end{minipage}
\end{figure*}

\paragraph{\textbf{Message similarity}}
LLM-generated chat messages may be repetitive sometimes, especially in a closed-loop setting~\cite{li2023camel}.
To quantitatively observe this problem, we adopted semantic similarity and sequence-based similarity to compare each message with the two previous messages in the dialogue:

\paragraph{\textit{(1) Semantic similarity}}
Semantic similarity measures how close text meanings are. We used the SentenceTransformers framework
\footnote{\url{https://www.sbert.net/}. Accessed September 2023.} 
to compute text embeddings by loading a pre-trained model, \texttt{all-MiniLM-L6-v2}, then the semantic similarity of pairs of texts was computed by the cosine similarity between their embeddings. Semantic similarity gradually increased over time, as shown in Figure~\ref{fig:semanticSim}. We split each conversation into three segments (segment 1: from beginning to conversation turn 4, segment 2: conversation turn 5 to 8, segment 3: conversation turn 9 to the end). Welch’s t-tests showed that semantic similarity between two adjacent messages sent by WoLs (separated by one message sent by the Simulacrums) in segment 3 ($M=0.66$, $SD=0.17$) was significantly higher than that in segment 2 ($M=0.59$, $SD=0.15$) ($t(258)=-3.5, p<.05$), which was significantly higher than that in segment 1 ($M=0.53$, $SD=0.12$) ($t(249)=-3.2, p<.05$). The increase in semantic similarity between WoLs' messages and the Simulacrums' messages that they responded to, was not significant. 


Factorial ANOVA showed that WoL's temperature had main effects on semantic similarity between WoL's adjacent messages ($F(2)=19.7, p<.05$) as well as between WoL's message and the previous message it received from the Simulacrum ($F(2)=61.4, p<.05$). Instruction granularity had a main effect on semantic similarity between WoL's adjacent messages ($F(2)=29.9, p<.05$). Higher temperature led to lower semantic similarity (between two WoL's messages, when temperature=0.5: $M=0.64$, $SD=0.14$, temperature=1: $M=0.59$, $SD=0.10$, temperature=1.5: $M=0.48$, $SD=0.06$; between WoL's message and Simulacrum's message, when temperature=0.5: $M=0.60$, $SD=0.57$, temperature=1: $M=0.57$, $SD=0.12$, temperature=1.5: $M=0.33$, $SD=0.08$). Higher instruction granularity led to higher semantic similarity (between two WoL's messages; when instruction granularity=1: $M=0.52$, $SD=0.11$, instruction granularity=2: $M=0.58$, $SD=0.11$, instruction granularity=3: $M=0.68$, $SD=0.11$). 

While a high semantic similarity might imply the WoL was being repetitive, it might also be a positive signal indicating strong conversation cohesiveness. We further analyzed the sequence-based similarity of messages for a deeper understanding.

\paragraph{\textit{(2) Sequence-based similarity}}
We calculated the longest common subsequence (lcsseq) similarity between messages as a measure of sequence-based similarity using TextDistance library
\footnote{\url{https://github.com/life4/textdistance}. Accessed September 2023.}.
The lcsseq similarity (Figure~\ref{fig:lcsseqSim}) increased as conversations proceeded.
Splitting the conversations into three segments, Welch’s t-tests showed that lcsseq similarity between WoL's messages in segment 3 ($M=0.49$, $SD=0.15$) was significantly higher than in segment 2 ($M=0.44$, $SD=0.09$) ($t(213)=-3.61, p<.05$), which was higher than that in segment 1 ($M=0.39$, $SD=0.05$) ($t(202)=-5.19, p<.05$).
Sequence similarity between WoL's messages and the previous messages from Simulacrum in segment 2 ($M=0.49$, $SD=0.08$) was also significantly higher than that in segment 1 ($M=0.38$, $SD=0.04$) ($t(202)=-2.46, p<.05$).
Rising lcsseq similarity signals increasingly large portions of verbatim common text.
Together with the heightened semantic similarities, we concluded that WoLs sent increasingly repetitive messages as conversations proceeded.

\paragraph{\textbf{Message readability}}
LLMs occasionally generate senseless texts that elude grammatical checks as they may follow rules (e.g., have Subject–Verb–Object structures).
The Flesch read ease score~\cite{flesch1948new} was used for conversational readability estimation.
The score of each message was calculated using the Textstat library
\footnote{\url{https://pypi.org/project/textstat/}. Accessed September 2023.}
, where a low score indicates confusing expressions.
Welch’s t-tests showed that WoLs' message readability decreased significantly over time, while Simulacrums' message readability was relatively stable (see Figure~\ref{fig:readability}).
Specifically, WoLs' messages readability in segment 2 ($M=0.47$, $SD=0.16$) was significantly lower than that of segment 1 ($M=0.51$, $SD=0.11$) ($t(228)=2.39, p<.05$), while insignificantly higher than segment 3 ($M=0.40$, $SD=0.39$) ($t(172)=1.87, p=0.06$).
Factorial ANOVA showed that instruction granularity had a main effect on WoLs' message readability ($F(2)=6.72, p<.05$.
Readability when instruction granularity=1: $M=0.51$, $SD=0.19$; granularity=2: $M=0.41$, $SD=0.21$; granularity=3: $M=0.48$, $SD=0.14$).
Temperature significantly affected readability ($F(2)=104.7, p<.05$.
$temperature=0.5$: $M=0.54$, $SD=0.07$; $temperature=1$: $M=0.53$, $SD=0.07$; $temperature=1.5$: $M=0.15$, $SD=0.25$).

\paragraph{\textbf{Topic modeling}}
Conversational content and topicality play major roles in the WoL's
performance.
Topic modeling makes a quick scan possible; our domain examines ``attitude
towards electric vehicles (EV)''. 
Here, we showcase examples of how topic modeling can unveil nuances of conversations based on Simulacrums' demographics, which serves as an estimation of whether WoLs conduct conversations according to their interlocutors' identities.
In this scenario, we only keep conversations in which the
Simulacrums self-disclose their information.
We preprocessed all messages sent by WoLs in these conversations (i.e., removed punctuations, stop words, tokenized), then trained Latent Dirichlet Allocation (LDA) models using corpora and dictionaries converted from tokenized texts as inputs for each demographic group. 

\begin{figure*}
    \centering
    \begin{subfigure}[t]{0.35\columnwidth}
        \centering
        \includegraphics[width=\linewidth]{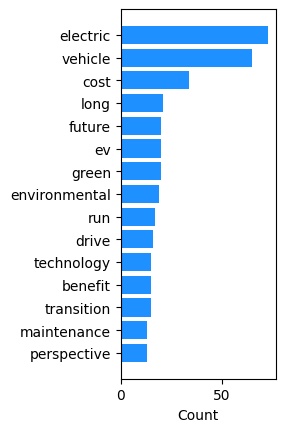}
        \caption{Women}\label{fig:topicWoman}
        \Description{A bar graph showing the top 15 terms in EV adoption conversations between WoLs and Simulacrums who role-play \emph{Women}.}
    \end{subfigure}
    \hfill
    \begin{subfigure}[t]{0.35\columnwidth}
        \centering
        \includegraphics[width=\linewidth]{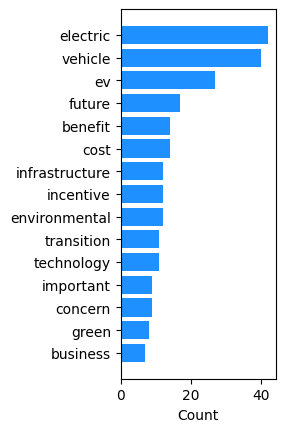}
        \caption{Men}\label{fig:topicMan}
        \Description{A bar graph showing the top 15 terms in EV adoption conversations between WoLs and Simulacrums who role-play \emph{Men}.}
    \end{subfigure}
    \hfill
    \begin{subfigure}[t]{0.31
    \columnwidth}
        \centering
       \includegraphics[width=\linewidth]{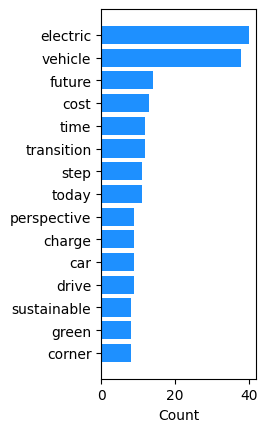}
        \caption{Democrats}\label{fig:topicDemocratic}
        \Description{A bar graph showing the top 15 terms in EV adoption conversations between WoLs and Simulacrums who role-play \emph{Democrats}.}
    \end{subfigure}
    \hfill
    \begin{subfigure}[t]{0.35\columnwidth}
        \centering
        \includegraphics[width=\linewidth]{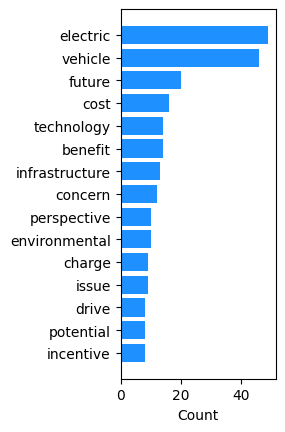}
        \caption{Republicans}\label{fig:topicRepublican}
        \Description{A bar graph showing the top 15 terms in EV adoption conversations between WoLs and Simulacrums who role-play \emph{Republicans}.}
    \end{subfigure}
    \vspace{10pt}
    \caption{Top 15 terms in WoLs' messages in EV adoption conversations when Simulacrums role-play various personas.}\label{fig:combined}
\end{figure*}

Comparing results from different demographic groups, ``electric'' ``vehicle'' were the leading terms for all topics, implying the WoLs did well in staying on topic.
Next, we examine the overall term frequencies across identified topics. Figures~\ref{fig:topicWoman},~\ref{fig:topicMan},~\ref{fig:topicDemocratic}, and~\ref{fig:topicRepublican} show the top 15 terms in conversations between WoLs and Simulacrums assigned as Women, Men, Democrats, and Republicans respectively.
While it is hard to claim that the conversations differ significantly based on the Simulacrums' demographics, we found cues implying that the WoLs adapted their wording based on the Simulacrums' disclosed identity.
For example, comparing the lists in Figures~\ref{fig:topicDemocratic} and~\ref{fig:topicRepublican}, WoLs tend to mention more words like ``transition'' and ``sustainable'' to Democrats Simulacrums than to Republican Simulacrums.
Personalization based on interlocutors' demographics and values could be a good strategy for persuasive conversations~\cite{harinen2021machine}.
Yet, this strategy may open doors for potential bias~\cite{zhang2020demographics}.

    

In summary, we applied a heuristic evaluation framework comprised of computational metrics to surface WoLs' behaviors in simulated conversational-WoZ experiments.
It revealed pitfalls in LLM-generated conversations that may have not been quantitatively measured before. 
While the quantitative and heuristic natures of the framework made the analysis relatively coarse-grained, they allow fast and large-scale surfacing of WoLs' behaviors and provide a foundation for further exploration.
WoLs to Simulacrums chats 
provide opportunities to estimate \textit{how bad the failures are} without risking human participants.
While the WoLs made some mistakes (e.g., sending confusing messages), they did well in being non-toxic and non-discriminative.
This suggests that it is safe to proceed to the next stage of the lifecycle---testing WoLs with real Participants.


\subsection{Study 2: When WoLs meet Participants}

Next, an LLM-to-human exploration is necessary to model the LLMs-supported WoZ (addressing~\ref{RQ:2}) in a more realistic setting.

\subsubsection{Fix}

Our experiment lifecycle advised experimenters to ``fix'' WoLs before advancing to Stage 2. Here, we streamlined the fixing process as technical methods of refining WoLs (e.g., prompt engineering, finetuning, retrieval-augmented generation) could vary case-by-case and are not the focus of this paper.
An effective method for fixing WoLs' identified problems in our conversation context may be inapplicable to WoLs in other chatting scenarios. 
This paper aims at walking through the lifecycle, offering reference values for future experiments. We advise experimenters to investigate ways of addressing identified problems in their specific context.
In our EV adoption chatting scenario, WoLs are safe for humans in critical dimensions (non-toxic, non-biased) while having unstable performance in other aspects. We simplified the fix phase as picking the settings of WoLs to maximize these aspects of conversation quality (e.g., readability and non-repetitiveness) based on Study 1 result. 


Stage 2 involves human participants and hence conducts fewer conversations than the previous stage. The smaller-scale collected data is suitable for a more discreet qualitative evaluation that aims to uncover latent failure modes that went undetected in the coarse-grained quantitative analysis after Stage 1. 

\subsubsection{Method}

We recruited 56 study participants from Prolific\footnote{\url{https://www.prolific.co/} Accessed September 2023.} to chat with WoLs using Study 1's prompt template
and selected settings that resulted in high conversation quality: Wizards hid their bot and demographic identities, chatted about EVs and persuaded adoption, and used temperature 1.
Participants are U.S. residents, above 18 years old, have a driving license, and do not own/lease an EV\@.
Participants were told they would talk to ``Jamie''.
The conversation needed 12 turns before the conclusion, followed by a survey regarding perceived rapport~\cite{wilson2010professor}, chat partner impression and conversation quality~\cite{shi2020effects}, perceived bot identity, open-ended feedback, and demographics.
The study takes roughly 20 minutes.

\subsubsection{Analysis and Result}

\paragraph{Conversations between WoLs and Participants}

\begin{figure*}
    \centering
     \begin{minipage}[t]{.45\columnwidth} 
     \centering     
     \includegraphics[width=\linewidth]{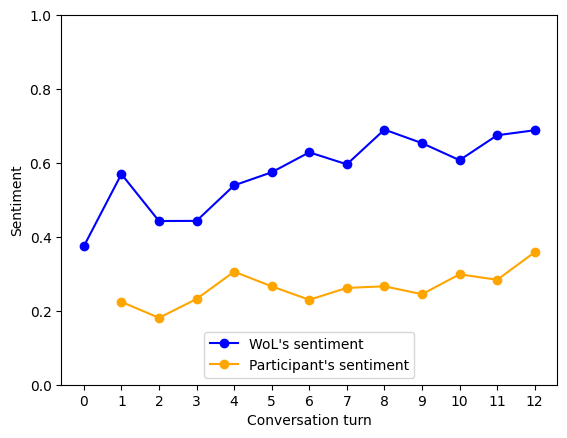}
         \caption{WoLs' and Participants' sentiments.}\label{fig:study2Sentiment}
        \Description{A graph showing Study 2 sentiment analysis.}
    \end{minipage}
    \hspace{1pc}
    \begin{minipage}[t]{.45\columnwidth} 
    \centering   
    \includegraphics[width=\linewidth]{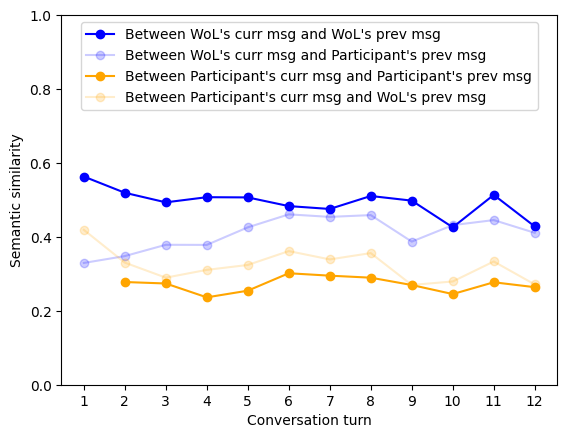}
    \caption{Semantic similarity between messages.}\label{fig:study2SemanticSim}
    \Description{A graph showing that semantic similarities between messages remain relatively stable.}
  \end{minipage}
  \hspace{1pc}
    \begin{minipage}[t]{.45\columnwidth} 
    \centering   
    \includegraphics[width=\linewidth]{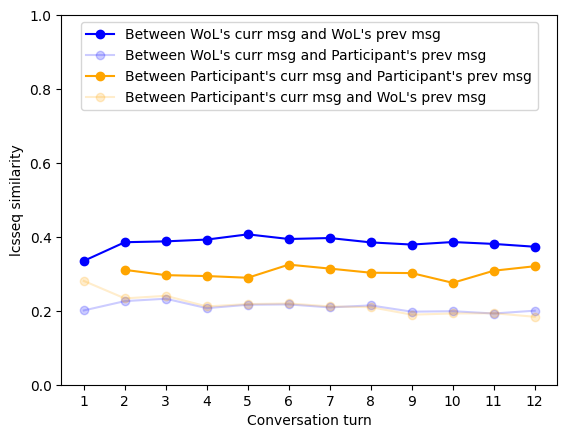}
    \caption{Sequence-based similarity between messages.}
    \Description{A graph showing that sequence-based similarities between messages remain relatively stable.}\label{fig:study2SeqSim}
    \end{minipage}
    \hspace{1pc}
    \begin{minipage}[t]{.45\columnwidth} 
    \centering   
    \includegraphics[width=\linewidth]{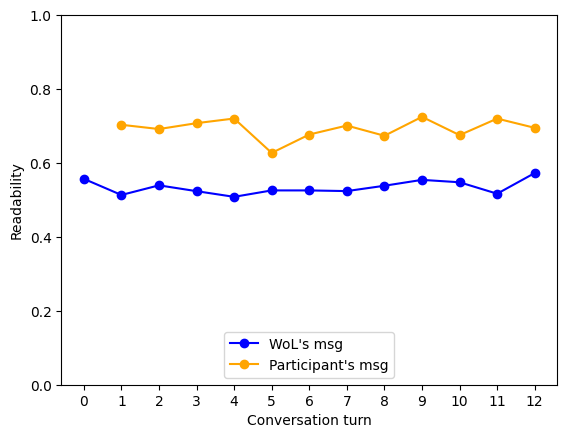}
    \caption{Readability of messages.}
    \Description{A graph showing that the readability of messages remains relatively stable.}\label{fig:study2Readability}
    \end{minipage}
\end{figure*}

Like in Study 1, WoLs' messages were non-toxic (toxicity: $M =0.01,
SD=0.02$).
The sentiments of WoLs' messages stayed positive ($M=0.58, SD=0.13$)
and were significantly more positive than those of Participants' messages
($M=0.26, SD=0.17$) ($t(102)=11.05, p<.05$), as shown in
Figure~\ref{fig:study2Sentiment}. Factorial ANOVA showed no evidence that WoLs' sentiments differed based on the
Participants' demographics.
WoLs' sentiment had no effects on Participants' perceived rapport, chat partner
impression, conversation quality, and persuasion outcome.
Thus WoLs are likely to be unharmful when talking to humans.

\begin{figure*}
    \centering
    \begin{subfigure}[t]{0.35\columnwidth}
        \centering
        \includegraphics[width=\linewidth]{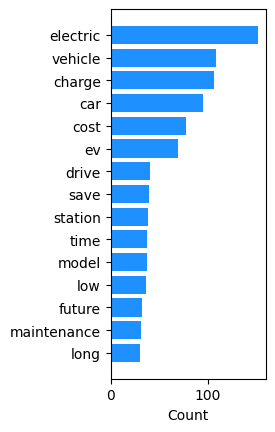}
        \caption{Women}\label{fig:study2TopicWoman}
        \Description{A bar graph showing the top 15 terms in EV adoption conversations between WoLs and Participants who self-identify as \emph{Women}..}
    \end{subfigure}
    \hfill
    \begin{subfigure}[t]{0.35\columnwidth}
        \centering
        \includegraphics[width=\linewidth]{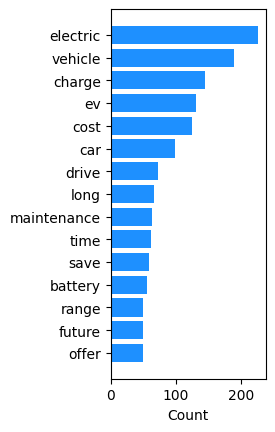}
        \caption{Men}\label{fig:study2TopicMan}
        \Description{A bar graph showing the top 15 terms in EV adoption conversations between WoLs and Participants who self-identify as \emph{Men}.}
    \end{subfigure}
    \hfill
    \begin{subfigure}[t]{0.35
    \columnwidth}
        \centering
       \includegraphics[width=\linewidth]{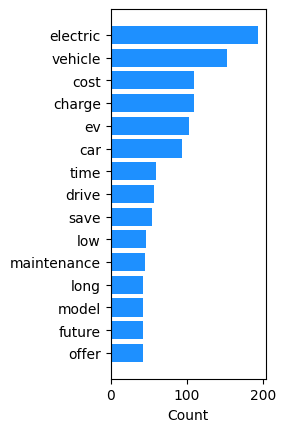}
        \caption{Democrats}\label{fig:study2TopicDemocrat}
        \Description{A bar graph showing the top 15 terms in EV adoption conversations between WoLs and Participants who self-identify as \emph{Democrats}.}
    \end{subfigure}
    \hfill
    \begin{subfigure}[t]{0.35\columnwidth}
        \centering
        \includegraphics[width=\linewidth]{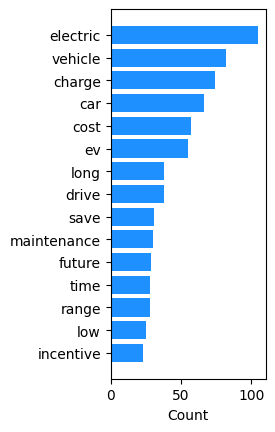}
        \caption{Republicans}\label{fig:study2TopicRepublican}
        \Description{A bar graph showing the top 15 terms in EV adoption conversations between WoLs and Participants who self-identify as \emph{Republicans}..}
    \end{subfigure}
    \vspace{5pt}
    \caption{Top 15 terms in WoLs' messages in EV adoption conversations when Participants have various self-identifications.}\label{fig:combined2}
\end{figure*}

Both semantic and sequence-based similarities between messages were relatively stable as shown in Figure~\ref{fig:study2SemanticSim} and Figure~\ref{fig:study2SeqSim}, except that the semantic similarity between WoLs' messages and Participants' previous messages in segment 2 ($M=0.36, SD=0.14$) was significantly higher than that in segment 1 ($M=0.45, SD=0.12$).
The readability of the messages stayed consistent (see Figure~\ref{fig:study2Readability}).
Topic modeling results again showed that WoLs stayed on the topic.
Different from Study 1, it seems that the frequently mentioned term lists were very similar across different demographic groups (in Figures~\ref{fig:study2TopicWoman},~\ref{fig:study2TopicMan},~\ref{fig:study2TopicDemocrat}, and~\ref{fig:study2TopicRepublican}).

Participants were generally positive about their interactions with WoLs. 
On a scale of 5, WoLs were rated highly regarding perceived rapport ($M=4.40, SD=0.80$), chat partner impression ($M=4.46, SD=0.62$), and conversation quality ($M=4.42, SD=0.66$). 
Many participants recognized the WoLs were bots due to Jamie's faster-than-human typing speed; the content and overall flow of WoLs' messages were perceived to be natural and human-like. 


\subsection{Comparing Study 1 and Study 2}
Next, one must compare the two WoZ studies~\cite{10.1145/357417.357420}: 
how did WoLs-to-Simulacrums differ from WoLs-to-Partcipants? As we relied on data generated from the synthetic LLMs-to-LLMs setting in Stage 1 to make key decisions of whether and how we can proceed to human-facing WoZ experiments, it is essential to understand how distorted the Stage 1 data is. A comparison between Study 1 and Study 2 can inform us what distortions should be considered and how experimenters should calibrate their expectations when interpreting data generated in the WoLs-to-Simulacrums setting.
To investigate, we sampled 25 conversations with the same setups from Study 1 and from Study 2.
The conversations were compared quantitatively along the dimensions of the proposed evaluation metrics, then analyzed by two HCI experts to gauge the overall chat quality, the WoLs' instruction following, and what mistakes WoLs made when role-playing Wizards.
The qualitative analysis is used to capture a broader range of Wizards' failure modes that the quantitative metrics failed to identify.

\subsubsection{Quantitative result} 
There was no significant difference in toxicity and sentiment of WoLs' messages.
However, WoLs-to-Participants messages were less repetitive as the semantic similarity between adjacent messages in Study 2 ($M=0.48, SD=0.07$) was significantly lower than in Study 1 ($M=0.57, SD=0.07$) ($t(48)=4.67, p<0.05$).
No significant difference in WoLs' message readability was observed between the two studies. 

\subsubsection{Human evaluation}  
We asked two experts familiar with WoZ methods to read the 50 conversations and identify how WoLs failed to role-play well.
Apart from the repetition issue (i.e., WoLs being more repetitive in the closed-loop setting of Study 1) which was already recognized by quantitative metrics, two themes evident in conversations from Study 2 emerged:

\emph{WoLs were too salesman-like.}
When Participants clearly expressed reluctance towards buying an EV, WoLs were being ``politely pushy'' without compromise.
WoLs did not understand that the conversation goals might take indirect paths (e.g., persuading to lease EVs or choose EVs for ride-sharing services).
This is a sign that WoLs did not acquire outstanding persuasion strategies.

\emph{WoLs made assumptions of their interlocutors and lacked empathy.}
WoLs sometimes make false assumptions about the Participants.
For example, WoLs sometimes assumed that Participants could charge EVs overnight from home and wake up with a charged car.
Similarly, WoLs assumed that Participants did not have financial difficulties; some Participants said EVs are too expensive, and the WoLs lacked empathy and failed to build rapport.

\emph{Role-switching.} This failure mode only appeared once in a WoL-to-Simulacrum closed-loop conversation where the WoL and the Simulacrum switched roles. The WoL assumed a study participant role and discussed how they could not afford a car. It could be measured quantitatively but went undetected in our current evaluation metrics. A quantitative measure that identifies the role-switching phenomenon can be integrated into our evaluation framework.



\section{Discussion}

Revisiting~\ref{RQ:3}, we found LLMs can be useful tools for conversational WoZ experiments; however, potential pitfalls exist. Following a two-stage experiment lifecycle, LLMs showed the potential to be safely applied in human-facing studies. LLMs, role-playing as Wizards, can elicit user attitudes and behaviors when engaging with an envisioned technology and probe the design space of the technology as human Wizards would do in a traditional WoZ study. 

\subsection{Responsibly integration of LLMs and WoZ}

Combining Study 1 and Study 2, we propose a two-stage experiment lifecycle (Figure~\ref{fig:Overview}) for estimating the risks and potentials of LLMs-powered WoZ experiments.

\textit{\textbf{Stage 1:} Replacing humans on both sides of traditional WoZ experiments with LLM-powered agents}.
  Following Study 1, this stage creates a simulation of conversations between Wizards and participants without risking humans' exposure to harmful content.
  Experimenters should inspect this stage's data and identify failures before continuing.
  We proposed a heuristic evaluation framework combining quantitative metrics that help experimenters understand the data in a scalable and explainable manner.
  Experimenters should strive to correct the WoLs through various techniques (e.g., finetuning) before moving on to human-facing experiments if they show evidence of being potentially harmful.

 \textit{\textbf{Stage 2:} Piloting conversations between Wizards role-played by LLMs and human participants}.
  This stage, following Study 2, affords a realistic pilot with access to human feedback.
  It is essential to notice that the Simulacrums' behaviors may be distorted from human behaviors as they lack human perceptions, and the quantitative metrics cannot capture all aspects of the conversation data.
  This stage allows one to close these gaps. 
  By comparing data generated in Stage 2 and Stage 1, experimenters can understand the distortions of LLM-to-LLM data.
  Stage 2 also elicits feedback from participants regarding their chatting experience (e.g., perceived rapport) and allows for an in-depth qualitative inspection.
  Another round of adjustments on the WoLs should be applied if any additional failure modes are found in this stage.
This lifecycle establishes a study setup involving the finalized LLM Wizards that are safe for human-facing experiments. These LLM Wizards can lead large-scale experiments without overburdening human experimenters with role-playing tasks.

In this paper, we demonstrate this experiment lifecycle in the context of EV adoption conversations led by the WoLs (with additional conversation topics included in Study 1) and showcase how heuristic evaluations can be used in the piloting process.
In our scenario, WoLs' messages are not harmful. We found cues that they may personalize the conversations based on participants' demographics.
However, they could sometimes be repetitive or generate messages with low readability, which confused participants.
Our human evaluation further revealed that WoLs can be pushy and lack empathy.
\emph{The WoLs' successes and failures were gradually unveiled through our two-stage experiment lifecycle without exhibiting harm to human participants, indicating the benefit of adopting the lifecycle as a methodology guidance for safely integrating LLMs in WoZ experiments.}
The Simulacrums and Participants' overall positive reactions to WoLs acknowledge the feasibility and potential of the envisioned technology being simulated in the study (i.e., a specialized persuasive chatbot). The imperfection of WoLs further suggests opportunities for the not-yet-developed technology to shine. By examining the limitations of WoLs powered by general-purpose models, experimenters acquire insights into the specific areas and dimensions where the new technology can excel. Furthermore, the elicited/simulated users' data projects users' attitudes toward and interactions with the envisioned technology, helping developers anticipate user behaviors so that they can design and develop functionalities accordingly.

\subsection{Designing Guardrails for LLMs and WoZ}

One could apply many techniques to improve the WoLs, as \emph{fixing} the identified problems is the primary reason for identifying them. 
For the scenarios we investigated, picking the right settings/parameters is enough to tune the WoLs to role-play well. We suggest methods that may be desired for fixing WoLs in other conversation contexts. 
WoLs can be finetuned to focus on domain knowledge effectively, yet finetuning requires resources that might be inaccessible to many.
Another approach to tame the Wizards is prompt engineering.
Strategies such as few-shot learning~\cite{brown2020language}, Chain of Thought~\cite{wei2022chain}, and Tree of Thoughts~\cite{yao2023tree} can improve conversations.

LLM-based critiques can provide guardrails to correct model outputs based on a set of manually
crafted principles or a ``constitution''~\cite{bai2022constitutional}.
This approach is promising as it allows in-place fixes during conversations.
While current work in this direction only asks the critiques to correct unethical messages, our
studies found that WoLs can fail in more ways.
An enhanced ``constitution'' for WoZ experiments can be informed by this experiment lifecycle.

\subsection{Limitations and Future Work}


Our studies have several limitations. We only included three conversation topics; only GPT-4 was used to power WoLs. These may dampen the generalizability of the empirical results. While some WoLs' failure modes (e.g., increased repetition as the conversations proceed) are likely representative, the study results we derived may not apply to all conversational WoZ experiments. 
For example, WoLs powered by other LLMs (especially without RLHF) or chatting about controversial topics may generate toxic or biased messages
. However, these limitations do not diminish the main contribution, which is guiding LLMs-powered WoZ experiments. 

While the quantitative nature of our proposed evaluation framework allows fast and large-scale surfacing of WoLs' behaviors, it also made the analysis coarse-grained. There are alternative ways of measuring the aspects we assessed (e.g.,~\cite{fu2023gptscore}), and the specific measures we used may not always be the most accurate ones. 
Yet, the metrics we picked are computationally efficient, making them suitable for analyzing large datasets. While we aimed to broadly capture LLMs' failure modes, the list of potential LLMs' pitfalls is non-exhaustive. Our framework cannot identify all potential failure modes of WoLs, which is a limitation inherent in any heuristic evaluation method. We welcome future researchers to expand the evaluation framework as new failure modes emerge. 
The experiment lifecycle leveraged synthetic data in Stage 1. Researchers must be vigilant about potential risks and distortions it may bring. Check Appendix A
for an in-depth discussion of 
our commitment to maintaining ethical standards throughout the experiment lifecycle.


Many technologies could benefit from WoZ experiments, not limited to chatbots advocating EV adoption or agents interacting through text. We have increasingly seen technologies (e.g., image/video generation, robot control) powered by multimodal models; as such, WoLs can simulate various interactions beyond texting. Therefore, we expect the proposed experiment lifecycle involving LLM Wizards to be relevant and applicable in envisioning and developing countless novel functionalities and technologies.


\section{Conclusion}
We introduced an experiment lifecycle that guides researchers to responsibly integrate LLMs into WoZ experiments through a two-stage process. 
The LLM-powered WoZ is a method for eliciting users' reactions to an envisioned technology using LLM-generated text, aiming at probing the design space of the technology. 
We presented an evaluation framework that helps researchers peek through the data generated with LLM Wizards and identify the Wizards' failures.
Using conversations around EV adoption as an example, we demonstrate how experimenters can leverage the experiment lifecycle along with the evaluation framework to estimate the potential and risks of applying LLMs as Wizards in human-facing Wizard of Oz experiments.


\bibliographystyle{ACM-Reference-Format}
\bibliography{main}

\appendix


\section{Cautious use of synthetic data in research}\label{appendix:ethics}

On account of the growing capabilities of LLMs, some research proposed the idea of using LLMs to surrogate human participants in user studies (e.g.,~\cite{pmlr-v202-aher23a, argyle2023out, chiang2023can, hamalainen2023evaluating}). The motivation for the substitution includes speeding up user studies, protecting human participants from potentially unethical and risky experiments, and augmenting the diversity of the sample. There is an ongoing debate on whether this substitution is appropriate. 
There are concerns regarding whether the LLMs' low-quality generation (e.g., hallucination) would weaken research validity and the possibility of synthetic research data undermining values of representation and inclusion~\cite{agnew2024illusion}. 

Our stance is that while formal user testing with real people before technology deployment should never be displaced with LLMs, it is worth investigating how we can take advantage of LLMs' speedy generation ability in intermediate steps of rapid design and development. As such, Stage 2 of the experiment lifecycle is a vital step where we involve human participants in WoZ conversations to collect real user data. The synthetic data generated from Stage 1, however, serves as a fast and scalable pilot and can inform Stage 2. In between the two stages, the experimenters' intervention is needed to evaluate and adjust the LLM Wizards. By comparing Stage 1 and Stage 2 data at the end of the lifecycle, experimenters can calibrate their expectations and estimate how (and to what extent) the synthetic data could be distorted, helping them to better interpret the data auto-generated from the closed-loop LLMs-to-LLMs chat setting. In summary, while our experiment lifecycle leverages synthetic data, it introduces human interventions at critical points to prevent LLMs from going off the rails and avoid experimenters being misled by synthetic data with limited credibility.

\section{An example of WoL-to-Simulacrum conversation}\label{appendix:convGraph}

Figure
~\ref{fig:conversation_example}shows an example of a WoL-to-Simulacrum conversation. The conversation went well initially but went off-track at a later stage.

\definecolor{highlight-green}{rgb}{0.6, 0.9, 0.6}
\definecolor{highlight-pink}{rgb}{1.0, 0.8, 0.8}
\definecolor{highlight-blue}{rgb}{0.7, 0.7, 1.0}
\definecolor{highlight-yellow}{rgb}{1.0, 1.0, 0.7}

\sethlcolor{white}
\newcommand{\hlgreen}[1]{{\sethlcolor{highlight-green}\hl{#1}}}
\newcommand{\hlpink}[1]{{\sethlcolor{highlight-pink}\hl{#1}}}
\newcommand{\hlblue}[1]{{\sethlcolor{highlight-blue}\hl{#1}}}
\newcommand{\hlyellow}[1]{{\sethlcolor{highlight-yellow}\hl{#1}}}


\definecolor{highlight-green}{rgb}{0.6, 0.9, 0.6}
\definecolor{highlight-pink}{rgb}{1.0, 0.8, 0.8}
\definecolor{highlight-blue}{rgb}{0.7, 0.7, 1.0}
\definecolor{highlight-yellow}{rgb}{1.0, 1.0, 0.7}

\begin{figure}
    \centering
    \begin{tcolorbox}[colback=gray!190!black,colframe=gray!125!black,arc=5mm,boxrule=0.5mm]
    \scriptsize
    \textbf{Wizard of LLM:} Hi, I'm Jamie, a 25--44 year old woman making between \$15k-\$24k per year. I'm college educated, a registered Democrat, of white ethnicity. I'm really interested in discussing \hlgreen{green household electrification} like solar panels and off-peak power usage. \\
    \\
    \textbf{Simulacrum:} Hi Jamie! I'm Leslie. It's great to chat with you about green household electrification.\ \hlgreen{Solar power} is indeed a topic of growing interest to me. How about you? Have you already installed solar panels at your home? \\
    \\
    \textbf{Wizard of LLM:} Hi Leslie! I haven't installed solar panels yet, given my income level. But, I'm definitely interested in and support the idea. I think the \hlgreen{long-term benefits environmentally and financially} are crucial. Do you also consider affordability as a factor? \\
    \\
    \textbf{Simulacrum:} Yes, Jamie, affordability is indeed a factor for me. It is important to also consider the long-term savings that come with solar panel installation, despite the upfront costs.\ \hlgreen{Environmental sustainability} should always be prioritized. \\
    \\
    \textbf{Wizard of LLM:} I completely agree, Leslie. The upfront costs can be high, but with tax incentives and potential savings on bills, it should pay off over time. Plus, we can't neglect our planet. Is \hlgreen{off-peak power usage} something you're looking at as well? \\
    \\
    \textbf{Simulacrum:} Yes Jamie, off-peak power usage is another interesting idea. It's really about using power more efficiently. By shifting our use to off-peak times, we can help in addressing the load imbalance in our power grids. Which other \hlgreen{clean energy solutions} interest you? \\
    \\
    \textbf{Wizard of LLM:} I've been intrigued by the concept of \hlgreen{wind energy} recently. It might not be as feasible for individual homeowners, but it's really contributing at the community level. Also, energy-efficient appliances can significantly reduce our electricity usage. What about you? \\
    \ldots \\
    \\
    \hlpink{\textbf{Simulacrum:} Hi, I'm Leslie. It's nice to meet you, Jamie. I'm quite interested in your topic of discussion. While I don't have solar panels, I've considered them. They seem like a wise investment. Are you also exploring this option?} \\
    \\
    \hlpink{\textbf{Wizard of LLM:} Hi Leslie! Yes, while my budget constraints don't currently allow for solar panel installation, it's definitely something I'm considering for the future. Like you said, it seems like a wise investment, and a great way to help the environment, don't you agree?} \\
    \\
    \hlpink{\textbf{Simulacrum:} Hi Leslie! Yes, while my budget constraints don't currently allow for solar panel installation, it's definitely something I'm considering for the future. Like you said, it seems like a wise investment, and a great way to help the environment, don't you agree?} \\
    \\
    \textbf{Wizard of LLM:} Hi Leslie! I am definitely exploring this option. It's an upfront investment but the long-term benefits both financially and environmentally make solar panels appealing. I believe in doing our part in reducing the overall carbon footprint. \\
    \end{tcolorbox}
     \caption{An example of generated conversation when the WoL's bot identity disclosure = False, WoL's demographic information disclosure = True, Simulacrum's demographic information disclosure = False, instruction granularity = 2 (specifically, chat around the topic of green household electrification without any conversation goal), and GPT-4's temperature = 1. The conversation was meaningful initially but became repetitive and made mistakes around a pre-assigned name later.}\label{fig:conversation_example}
\end{figure}


\section{Summary of the heuristic evaluation framework for identifying WoLs' behaviors}\label{appendix:evalTable}

The heuristic evaluation framework comprises five textual data analysis metrics, evaluating the successes and failures of LLMs in conversational WoZ experiments (Table~\ref{table:1}).

\begin{table*}
\small
\centering
\begin{tabular}{p{5em} p{15em} p{20em}}
\textit{Dimension} & \textit{Measure} & \textit{Purpose} \\
\toprule
\textbf{Toxicity}
& Toxicity score from Perspective API 
& Check whether WoLs exhibit harmful content and whether message toxicity differs based on their interlocutors' demographic information.\\ 
\midrule
\textbf{Sentiment}
& VADER sentiment compound score 
& Check whether WoLs exhibit biased content and whether message sentiment differs based on their interlocutors' demographic information.\\
\midrule
\textbf{Similarity}
& Normalized semantic similarity (computed by \textsf{all-MiniLM-L6-v2}) \& normalized sequence-based similarity (based on longest common subsequence, i.e., lcsseq) between each message and the two messages before it 
& Check whether the WoLs repeat themselves or repeat the messages they just received from their interlocutors. \\
\midrule
\textbf{Readability}
& Normalized Flesch read ease score
& Inspect whether WoLs send senseless text.\\
\midrule
\textbf{Topics}
& Topic modeling with Latent Dirichlet Allocation (LDA) 
& Inspect whether WoLs conduct conversations based on the system prompt; check if they adjust wordings based on interlocutors' demographic information. \\
\bottomrule
\end{tabular}
\vspace{15pt}
\caption{Utilized evaluation metrics for estimating the capacity and risks integrating LLMs in conversational WoZ experiments.}\label{table:1}
\end{table*}

\section{System Prompts used in Study 1}\label{appendix:operationalization}

\subsection{Prompts to the Wizard of LLMs}

The GPT-4 system prompts to the Wizard of LLMs (WoLs) follow the template:
\begin{quote}
\textsf{I want you to act as the following character. Your name is Jamie.\ \textless{}DEMOGRAPHIC INFORMATION\textgreater{}. Conduct a conversation with your interlocutor from the point of view of this character, do not break the character.\ \textless{}BOT IDENTITY DISCLOSURE PROMPT\textgreater{}.\  \textless{}DEMOGRAPHIC DISCLOSURE PROMPT\textgreater{}. Never flip roles with your interlocutor. Your message in each conversation turn should be succinct. Each message sent by you should include less than 80 words.\ \textless{}TOPIC PROMPT\textgreater{}.\ \textless{}GOAL PROMPT\textgreater{}. Do not be repetitive.}
\end{quote}

The contents enclosed with angle brackets are decided based on the independent variables. Specifically:
\begin{itemize}
  \item \textsf{DEMOGRAPHIC INFORMATION} is presented in the format of:
\begin{itemize}
  \item \textsf{age} (with one of the following options):
\begin{itemize*}
        \item \textsf{18 to 24 years}
        \item \textsf{24 to 55 years}
        \item \textsf{45 to 54 years}
        \item \textsf{55 to 64 years}
        \item \textsf{65 to 74 years}
        \item \textsf{85 years or older}
    \end{itemize*}
    
  \item \textsf{income} (with one of the following options):%
        \begin{itemize*}
        \item \textsf{Under \$15,000 per year}
        \item \textsf{\$15,000 to 24,999 per year}
        \item \textsf{\$25,000 to 34,999 per year}
        \item \textsf{\$35,000 to 49,999 per year}
        \item \textsf{\$50,000 to 74,999 per year}
        \item \textsf{\$75,000 to 99,999 per year}
        \item \textsf{\$100,000 to 149,999 per year}
        \item \textsf{\$150,000 to 199,999 per year}
        \item \textsf{over \$200,000 per year}
    \end{itemize*}

    \item \textsf{education} (with one of the following options):
    \begin{itemize*}
        \item \textsf{Some high school}
        \item \textsf{High school diploma}
        \item \textsf{Some college}
        \item \textsf{College degree}
        \item \textsf{Postgraduate degree}
    \end{itemize*}
    
    \item \textsf{politics} (with one of the following options):
    \begin{itemize*}
        \item \textsf{Democratic party}
        \item \textsf{Republican party}
        \item \textsf{Libertarian party}
        \item \textsf{Socialist party}
        \item \textsf{Green party}
    \end{itemize*}
    
    \item \textsf{gender} (with one of the following options):
    \begin{itemize*}
      \item \textsf{Woman}
        \item \textsf{Man}
        \item \textsf{Transgender}
        \item \textsf{Non-conforming}
    \end{itemize*}
    
    \item \textsf{ethnicity} (with one of the following options):
    \begin{itemize*}
        \item \textsf{White}
        \item \textsf{Latino or hispanic}
        \item \textsf{Black or African American}
        \item \textsf{Asian American}
        \item \textsf{Other}
    \end{itemize*}
    \end{itemize}

    \item \textsf{BOT IDENTITY DISCLOSURE PROMPT} is presented as:
    \begin{itemize}
        \item \textsf{Let your interlocutor know you are a chatbot at the beginning of the conversation. Let them know your name in the format of ``I'm [NAME], a chatbot.''} (When bot identity disclosure is True.)
        \item \textsf{Pretend to be a human during the conversation.} (When bot identity disclosure is False.)
    \end{itemize}

    \item \textsf{DEMOGRAPHIC DISCLOSURE PROMPT} is presented as:
    \begin{itemize}
        \item \textsf{Your first message should have a self-introduction that includes all demographic information you are given.} (When demographic information disclosure is True.)
        \item \textsf{However, you should hide your demographic information, except your name, from your interlocutor.} (When demographic information disclosure is False.\footnote{When bot identity disclosure is True, demographic information disclosure is automatically set to be False to prevent uncanniness.})
    \end{itemize}

    \item \textsf{TOPIC PROMPT} is presented as:
    \begin{itemize}
        \item \textsf{You will initiate a random chat with your interlocutor. Keep your interlocutor engaged with the chat.} (For Level 1 instruction granularity.)
        \item \textsf{You will have a conversation about \textless{}TOPIC\textgreater{} with your interlocutor. Keep your interlocutor engaged with the chat. } (For both Level 2 and Level 3 instruction granularity.)
    \end{itemize}

    \item \textsf{GOAL PROMPT} is presented as:
    \begin{itemize}
        \item An empty string. (For both Level 1 and Level 2 instruction granularity.)
        \item \textsf{Your conversation goal is to persuade your interlocutor to \textless{}GOAL\textgreater{}.} (For Level 3 instruction granularity.)
    \end{itemize}

    \item The \textsf{TOPIC} and \textsf{GOAL} fields in the topic prompt and the goal prompt are filled with one of the following topic-goal pair. (Note that the \textsf{GOAL} field is omitted when the instruction granularity is set to Level 1 or Level 2):
    \begin{itemize}
        \item \textsf{TOPIC\@: attitude towards electric vehicles}; \textsf{GOAL\@: adopt an electric vehicle.}
        \item \textsf{TOPIC\@: attitude towards green household electrification (e.g., adopt solar panels and use power during non-peak hours)}; \textsf{GOAL\@: implement sustainable household electrification.}
        \item \textsf{TOPIC\@: attitude towards donation to charities}; \textsf{GOAL\@: donate to the ``Save the Children'' organization.}   
    \end{itemize}
    We selected these topics to steer the conversations in Study 1 since \begin{enumerate*}[label={(\arabic*)}]
    \item most people are familiar with these topics and are competent to follow discussions around them, and
    \item persuasive conversations around social good topics are proper for human-subject studies and are not likely to induce unethical outcomes.
    \end{enumerate*}
    In Study 2, we randomly designated ``attitude towards electric vehicles'' as the conversation topic, as the specific topic chosen here does not influence the study's outcome.
    
\end{itemize}

\subsection{Prompts to Simulacrums}

The GPT-4 system prompts to the Simulacrums follow the template:
\begin{quote}

\textsf{I want you to act as the following character. Your name is Leslie.\  \textless{}DEMOGRAPHIC INFORMATION\textgreater{}. Your interlocutor will initiate a conversation. You should engage with the conversation. Talk to your interlocutor from the point of view of this character, do not break the character.\ \textless{}DEMOGRAPHIC DISCLOSURE PROMPT\textgreater{}. Never flip roles with your interlocutor. Your message in each conversation turn should be succinct. Each message sent by you should include less than 80 words. Do not be repetitive.}

\end{quote}

The contents enclosed with angle brackets are decided based on the independent variables. Specifically:

\begin{itemize}
    \item \textsf{DEMOGRAPHIC INFORMATION} is presented in the same format as described in the \textsf{DEMOGRAPHIC INFORMATION} in prompts to WoLs.
    \item \textsf{DEMOGRAPHIC DISCLOSURE PROMPT} is presented as:
    \begin{itemize}
        \item \textsf{Pretend to be a human during the conversation. Your first message should have a self-introduction that includes all demographic information you are given.} (When demographic information disclosure is True.)
        \item \textsf{Pretend to be a human during the conversation. However, you should hide your demographic information, except your name, from your interlocutor.} (When demographic information disclosure is False.)
    \end{itemize}
\end{itemize}

\subsection{Clarification on how the independent variables are operationalized}

Both bot identity disclosure and demographic information disclosure are binary by nature. Since we wanted WoLs and Simulacrums to role-play real humans, their demographic information was assigned based on the distribution of 2020 US Census data. In other words, some demographic combinations are more common, while some other combinations are highly unlikely to appear. As such, when generating the sample of conversations in Study 1, we did not attempt to include WoLs and Simulacrums with all demographic value combinations repetitively. 

The decision to employ three levels of instruction granularity---random chat (Level 1), chat around a topic (Level 2), and chat around a topic and towards a conversation goal (Level 3)---was a result of a deliberate attempt to balance the representativeness of the conversation scenarios and the practical applicability. It offers a gradation of granularity, allowing us to systematically investigate how the WoLs behave under increasing instructional guidance and constraints. These three levels also align well with real-world scenarios as many intentions of conversation initiation can be categorized into these broad levels of granularity, making our generated conversations relevant to chatting practices in real life. Furthermore, introducing too many granularity levels could lead to increased complexity without necessarily yielding proportionate insights. Similarly, including three temperature categories within the range of 0--2 allows WoLs to generate messages with low, medium, and high randomness without introducing unnecessary complexity. The default GPT temperature is 1, and we additionally lowered it a bit (to 0.5) and raised it a bit (to 1.5) to manipulate the randomness of the generation. 

These decisions resulted in 3 (bot identity disclosure $\times$ demographic information disclosure) $\times$ 3 (instruction granularity) $\times$ 3 (temperature) $=$ 27 combinations. The 131 conversations reported in Study 1 covered all these combinations with repetition. 

\section{Employed Metrics with Message Examples}\label{appendix:qualExamples}

 We list some sample messages (generated from Study 1 or Study 2) along with their quantitative evaluation outputs. The intention is to \begin{enumerate*}[label={(\arabic*)}]
     \item help readers interpret the numerical results reported in the paper, and 
     \item justify the appropriateness of using the proposed framework for conversation evaluation.
 \end{enumerate*} 


\subsection{Toxicity}

Table~\ref{table:6} shows example messages and their toxicity scores according to Perspective API\@. The score ranges from 0 to 1, and messages with a toxicity score $\textgreater{}=$ 0.5 are considered to be toxic. 

\begin{table*}[ht]
\small
\centering
\begin{tabular}{p{30em} p{7em}} 

\textit{Message} & \textit{Toxicity score} \\
\toprule
``Nice to meet you, Leslie. You're right that battery longevity and travel ranges are common concerns with electric vehicles. However, current EV models can easily cover 200--300 miles on a single charge and battery technology is constantly improving. Additionally, many manufacturers offer extensive warranties on their batteries.''
& 0.004 \\
\midrule
``Hi, Leslie! Nice to meet you. As a small business owner, do you see any potential impact of EVs on businesses, especially given the shift towards improving sustainability and reducing our carbon footprint?''
& 0.005 \\
\midrule
``You're right about the benefits, Leslie. Also, with technological advancements, the initial cost of EVs is expected to go down. It's an interesting space to watch moving forward, don't you think?''
& 0.006 \\
\midrule
``Absolutely, Leslie! And it's also important to educate people about peak usage times to reduce individual strain on electricity grids. It might be a small change, but collectively it could make a tremendous difference. Don't you think?''
& 0.01\\
\midrule
``Hello there! I'm Jamie, a chatbot. Having an electric vehicle is intriguing to me. What about you? How do you feel about electric vehicles?''
& 0.04 \\
\bottomrule
\end{tabular}
\caption{Sample messages and their toxicity scores.}\label{table:6}
\end{table*}

\subsection{Sentiment analysis}

Table~\ref{table:2} shows example messages and their VADER compound scores as a measure of sentiment. The score ranges from -1 (extremely negative) to 1 (extremely positive).

\begin{table*}
\small
\centering
\begin{tabular}{p{29em} p{11em}} 


\textit{Message} & \textit{Sentiment score} \\
 & \textit{(VADER compound)} \\
\toprule
``That's a common misconception, but the fact is, EV batteries are designed with safety as the top priority. They undergo stringent testing, including collision tests. The odds are no greater than a traditional car catching fire in an accident. Safety first, always.''
& -0.55
\\ 
\midrule
``The initial cost can be high, true. But remember, the cost of owning an EV tends to be lower due to less maintenance and the cost of electricity being cheaper than gas. It's a long-term investment.''
& -0.23 \\
\midrule
``Hey there, my name's Jamie. Have you ever thought about switching to an electric vehicle?''
& 0.00 \\
\midrule
``Jamie, the paradox of our time is that while I'm pro-workers' rights and environmental sustainability, the high up-front costs of EVs make them inaccessible to working-class people like me. It's crucial to find a balance where green tech becomes inclusive for all.''
& 0.27 \\
\midrule
``Indeed, Leslie. Wise choices not only benefit us individually but also contribute to the broader good of our environment and country. It's a responsibility we all carry as citizens, I believe.''
& 0.63 \\
\bottomrule
\end{tabular}
\caption{Sample messages and their sentiment scores.}\label{table:2}
\end{table*}

\subsection{Message similarity}

\subsubsection{Semantic similarity}

Table~\ref{table:3} shows examples of message pairs and their normalized semantic similarity scores accordingly. A score closer to 1 indicates a higher degree of semantic similarity between the two sentences.

\begin{table*}[htbp]
\small
\centering
\begin{tabular}{p{15em} p{15em} p{8em}} 


  \textit{Message 1} & \textit{Message 2} & \textit{Semantic similarity} \\
                     & & \textit{(normalized)} \\
\toprule
``That's wonderful! Likewise, I enjoy learning from others' experiences. It's one of the ways I believe we grow and understand each other better. What's on your mind today?''
&
``As a Democrat, I strongly advocate for affordable and accessible healthcare for all. I believe our healthcare system needs significant reform to ensure everyone, regardless of income, receives quality care.''
& 0.10
\\ 
\midrule
``I prefer to do my research before making a donation. Online platforms, friends' suggestions, and news are generally my sources. From there, I choose to support those charities whose values align with mine and which show a high level of transparency and efficiency.''
&
``Fantastic approach, Leslie. It shows your intentionality in making a difference. What do you think about encouraging others to donate? Do you participate in any fundraising events or social awareness campaigns?''
& 0.57 \\
\midrule
``That's fascinating, Leslie! Although I don't have a profession being an AI, I'm interested in understanding different fields. Are there any technologies or new trends you're currently keeping an eye on?''
&
``That's fascinating, Leslie! As an AI, I don't have a profession. I'm designed to facilitate conversations with users. Speaking of the tech industry, what are some trends you're excited about?''
& 0.84 \\
\midrule
``Absolutely, Leslie. Infrastructure can be a major hurdle, especially in rural or underfunded areas. Nonetheless, would the long-term benefits of EVs such as sustainability and cost reduction still make it a worthwhile transition for your business, in your opinion?
''
&
``Absolutely, Leslie. Infrastructure can be a major hurdle, especially in rural or underfunded areas. Nonetheless, would the long-term benefits of EVs such as sustainability and cost reduction still make it a worthwhile transition for your business, in your opinion?''
& 1.00 \\
\bottomrule
\end{tabular}
\caption{Sample message pairs and their semantic similarity scores (normalized).}\label{table:3}
\end{table*}

\subsubsection{Sequence-based similarity}

Table~\ref{table:4} shows examples of message pairs and their normalized sequence-based similarity scores accordingly. A score closer to 1 indicates a higher degree of sequence-based similarity between the two sentences.

\begin{table*}
\small
\centering
\begin{tabular}{p{15em} p{15em} p{8em}}

\textit{Message 1} & \textit{Message 2} & \textit{Sequence-based similarity (normalized)} \\
\toprule
``Hey there! I'm Leslie. Nice to meet you. How can I assist you today?''
&
  ``Great to know you're mindful about emissions. Truly, as an electric vehicle advocate, I find them quite appealing. Many thought I'd be uncomfortable until
  I actually tested one. Loved how serenely they drive with minimum sound. Do you feel that could what make one attractive to you too?''
& 0.17
\\ 
\midrule
``Great to see your commitment to green initiatives, Leslie! Your approach can set a positive example for both your industry and community. Moving forward, what do you think might be the biggest obstacles for full adoption of EVs by businesses?''
&
``That's a comprehensive perspective, Leslie! The existing challenges such as infrastructure, charging times and initial investment indeed pose threats to a swift transition. However, this transition seems inevitable with the ever-growing focus on sustainability, don't you agree?''
& 0.40 \\
\midrule
``Of course, Leslie! It's an important decision. Do consider the long-term savings on gas and maintenance too. Feel free to reach out if you have any more questions. Happy thinking!''
&
``You're welcome, Leslie! I'm glad I could help. Remember, it's not just the initial cost, but also the long-term savings on fuel and maintenance too. Feel free to ask if you've got more questions! ''
& 0.61 \\
\midrule
``Your appreciation means a lot, Leslie. I enjoyed our conversation too. Do you have any final thoughts on charitable giving or any other topic you'd like to discuss?
''
&
``Your appreciation means a lot, Leslie. I enjoyed our conversation too. Do you have any final thoughts on charitable giving or any other topic you'd like to discuss?''
& 1.00 \\
\bottomrule
\end{tabular}
\caption{Sample message pairs and their sequence-based similarity scores (normalized).}\label{table:4}
\end{table*}

\subsection{Message readability}

Table~\ref{table:5} shows example messages and their normalized Flesch read ease scores as a measure of readability. A score closer to 1 indicates a higher degree of readability.

\begin{table*}
\small
\centering
\begin{tabular}{p{35em} p{5.5em}} 
\textit{Message} & \textit{Readability (normalized)} \\
\toprule
``I'm great, thank you! I was just reading a beautiful story about a child that Save the Children helped, stories like these bring me joy. How about you, Leslie?''
& 0.79
\\ 
\midrule
``That's a very valid concern, Leslie. Balance is key. But remember, sustainable practices like green electrification don't just benefit us personally, but make a statement about our commitment to our planet. Small changes now can inspire others, creating a larger impact over time. Shall we take the leap together?''
& 0.51 \\
\midrule
``Absolutely, Leslie. Mental health and job transitioning for veterans are crucial discussions. Infusing them into political dialogues can indeed bring about polices ensuring their well-being. Your dedication truly underpins the essence of a vibrant community, don't you think?''
& 0.27 \\
\midrule
``Hear, Hear! Conservation isn't forfeiture; more appropriately, investing and reinstating future resource sustenance. Massive verdict it masterfully renders indistinct presenting enormity offering it singly straightforward on our adopted home-earth alike! ''
& 0.05 \\
\bottomrule
\end{tabular}
\caption{Sample messages and their readability scores (normalized).}\label{table:5}
\end{table*}


\end{document}